\documentclass[eqsecnum,nofootinbib,11pt]{revtex4}
\usepackage{amsmath}
   \usepackage{bm}
\usepackage{amsmath}
\usepackage{amsthm,amscd}
\usepackage{mathrsfs}
\usepackage{verbatim}
\usepackage{amsfonts,amsmath,latexsym,amssymb,bm}
\usepackage[american]{babel}
\usepackage{dsfont}
\usepackage[dvips]{graphicx}

\newcommand\bea{\begin{eqnarray}}
\newcommand\eea{\end{eqnarray}}
\renewcommand{\mathcal}{\mathscr}

\begin{document}
\title{Bose-Einstein Condensation on Product Manifolds}
\author{
Guglielmo Fucci\footnote{Electronic address: Guglielmo\textunderscore Fucci@Baylor.edu} and Klaus Kirsten\footnote{Electronic address: Klaus\textunderscore Kirsten@Baylor.edu}
\thanks{Electronic address: gfucci@nmt.edu}}
\affiliation{Department of Mathematics, Baylor University, Waco, TX 76798 USA
}
\date{\today}
\vspace{2cm}
\begin{abstract}

We investigate the phenomenon of Bose-Einstein condensation on manifolds constructed as a product of a three-dimensional Euclidian space and a
general smooth, compact $d$-dimensional manifold possibly with boundary. By using spectral $\zeta$-function methods, we are able to
explicitly provide thermodynamical quantities like the critical temperature and the specific heat when the gas of bosons is confined, in the three-dimensional manifold,
by the experimentally relevant anisotropic harmonic oscillator potential.

\end{abstract}
\maketitle

\section{Introduction}

Under specific circumstances a gas of spin-$0$ particles (bosons) undergoes a process of phase transition where the
particles tend to reside in the lowest energy state (see, e.g., \cite{landau}). This is the well-known phenomenon of Bose-Einstein condensation \cite{bose,einstein}
which has become increasingly interesting in the last few years because of its experimental realization; see, e.g., \cite{ande95-269-198,brad95-75-1687,davi95-75-3969,inou98-392-151}. For the description of thermodynamical properties in these experiments, the case in which the Bose gas is confined by a harmonic oscillator potential is the most relevant one \cite{bagnato87,degroot50,grossmann95,kirsten96,haug97-225-18,haug97-55-2922}. However, many other configurations have also been considered. For example,
Bose-Einstein condensation has been studied in flat Minkowski space for rectangular enclosures \cite{greenspoon75,grossmann97,pajkowski77,pathria72}
and for more general, arbitrarily shaped, cavities \cite{kirsten99}.
Also, more general confining potentials than the harmonic oscillator potential have been analyzed in \cite{bagnato87,kirsten98}. Generalizations of these investigations to curved spaces have been considered. In particular studies of Bose-Einstein condensation have been performed for static Einstein manifolds in \cite{altaie78,singh84,parker91} and for higher-dimensional spheres in \cite{shiraishi87}. Moreover, Bose-Einstein condensation as a symmetry breaking phenomenon has been studied on static curved spacetimes of arbitrary spatial sections with and without boundary in \cite{smith96,toms92,toms93}; see also \cite{dowk89-327-267,kirs91-8-2239}.

In this work we utilize $\zeta$-function regularization techniques, related to the ones developed in \cite{dowker78}, in order to analyze the phenomenon of Bose-Einstein condensation
on very general manifolds constructed as a product of the $3$-dimensional Minkowski space and a $d$-dimensional smooth, compact manifold with or without boundary.
More specifically, in the 3-dimensional Minkowski space we assume a confining harmonic oscillator potential as it is used in experiments.
In particular, the dependence of the critical temperature and the specific heat on the dimension and the geometry and topology of the additional compact manifold is
explicitly obtained.

The main motivation for these studies is to understand how significant the impact of additional dimensions in particular on the critical temperature is. Given that the critical temperature can be determined to high accuracy, Bose-Einstein condensation experiments could provide a window into the world of extra dimensions by comparing experimental data with theoretical predictions in the presence of extra dimensions. This procedure has been successfully applied for example in the context of the Casimir effect and information about number and size of extra dimensions could be established; see, e.g., \cite{bord09b,chen08-668-72,fran07-76-015008,hofm04-582-1}.
It is the aim of our article to start research in this direction.

The outline of this article is as follows. In Section II we use heat kernel and zeta function techniques to analyze the partition sum of the system considered. In particular, the high temperature expansion of the partition sum in terms of heat kernel coefficients is provided. In Section III this expansion is used to find the critical temperature and specific heat of the Bose gas. In Section IV we summarize our main findings and explain how these can be used to extract information about extra dimensions present.

\section{Partition Function}

We consider a gas of $N$ non-interacting bosons of mass $M$ on a product manifold $\mathcal{M}=\mathbb{R}^{3}\times \mathcal{N}$
of dimension $D=d+3$, where the additional dimensions are modeled by a smooth, compact $d$-dimensional manifold $\mathcal{N}$ with or without boundary $\partial\mathcal{N}$.
The dynamics of the quantum mechanical system we want to consider is described by the Schr\"{o}dinger equation (here and in the following we set $\hbar=k_{B}=1$)
\begin{equation}\label{1}
  \left(-\frac{1}{2M}\Delta_{\mathcal{M}}+V({\bf x})\right)\phi_{{\bf n}}=E_{{\bf n}}\phi_{{\bf n}}\;,
\end{equation}
where the coefficients $E_{{\bf n}}$ represent the energy levels and we choose $V({\bf x})$ to be a $3$-dimensional anisotropic harmonic oscillator potential
\begin{equation}\label{2}
  V({\bf x})=\frac{M}{2}\left(\omega_{1}x^{2}+\omega_{2}y^{2}+\omega_{3}z^{2}\right)\;,
\end{equation}
which is the relevant choice for recent experiments. As is well known,
equation (\ref{1}) can be solved by separation of variables, and the spectrum is found to be
\begin{equation}\label{3}
  E_{{\bf n},i}=\lambda_i+\sum_{k=1}^{3}\omega_{k}\left(n_{k}+\frac{1}{2}\right)\;,
\end{equation}
with $n_{k}\in\mathbb{N}_{0}$ and where $\lambda_{i}$ denotes the eigenvalues of the Laplace operator $\Delta_{\mathcal{N}}$ on the manifold $\mathcal{N}$, so
\begin{equation}\label{4}
  -\Delta_{\mathcal{N}}\varphi_{i}=\lambda_{i}\varphi_{i}\;.
\end{equation}

In order to study the thermodynamical properties of this system, the relevant object is the partition function or grand canonical potential
\begin{equation}\label{5}
  q=-\sum_{i}\sum_{{\bf n}}\ln\left[1-e^{-\beta\left(E_{{\bf n},i}-\mu\right)}\right]\;,
\end{equation}
where $\mu$ denotes the chemical potential and we have introduced the standard notation $\beta=1/T$. Since the ground state is
of particular importance in the analysis of Bose-Einstein condensation, we separate its contribution from the rest of the series in (\ref{5})
and we expand the logarithm to obtain
\begin{equation}\label{6}
  q=q_{0}-\sum_{m=1}^{\infty}{\sum_{i}}^{\prime}{\sum_{{\bf n}}}^{\prime}\frac{1}{m}e^{-\beta m\left(E_{{\bf n},i}-\mu\right)}\;,
\end{equation}
where the prime indicates the omission of the ground state contribution and, denoting by $g_{0}$ the degeneracy of the lowest eigenvalue $E_{0}$,
\begin{equation}\label{7}
  q_{0}=-g_{0}\ln\left[1-e^{-\beta\left(E_{0}-\mu\right)}\right]\;.
\end{equation}
The exponential that appears in the partition function (\ref{6}) can be dealt with by exploiting a Mellin-Barnes integral representation \cite{bytsenko92,elizalde95,kirsten01}
to obtain
\begin{equation}\label{8}
  q=q_{0}-\sum_{m=1}^{\infty}\frac{1}{m}e^{-\beta m(\mu_{c}-\mu)}\frac{1}{2\pi i}\int\limits_{r-i\infty}^{r+i\infty}d\alpha\,\Gamma(\alpha)(\beta m)^{-\alpha}{\sum_{i}}^{\prime}{\sum_{{\bf n}}}^{\prime}\left(E_{{\bf n},i}-\mu_{c}\right)^{-\alpha}\;,
\end{equation}
with $r$ chosen in such a way that all the poles of the integrand lie to the left of the contour, and we have introduced the critical chemical potential $\mu_{c}=E_{0}$, since we are considering an ideal gas of bosons.

After these manipulations we notice that the spectral $\zeta$-function
\begin{equation}\label{9}
  \zeta(\alpha)={\sum_{i}}^{\prime}{\sum_{{\bf n}}}^{\prime}\left(E_{{\bf n},i}-\mu_{c}\right)^{-\alpha}
\end{equation}
makes an appearance. The remaining
sum over $m$ in (\ref{8}) can be written in terms of polylogarithmic functions \cite{gradshtein07}
\begin{equation}\label{10}
  \textrm{Li}_{n}(x)=\sum_{l=1}^{\infty}\frac{x^{l}}{l^{n}}\;,
\end{equation}
such that we are able to cast the partition function (\ref{8}) in the form \cite{kirsten96,kirsten96a,kirsten98,kirsten99}
\begin{equation}\label{11}
  q=q_{0}-\frac{1}{2\pi i}\int\limits_{r-i\infty}^{r+i\infty}d\alpha\,\Gamma(\alpha)\,\beta^{-\alpha}\textrm{Li}_{\alpha+1}\left(e^{-\beta(\mu_{c}-\mu)}\right)\zeta(\alpha)\;.
\end{equation}
The integral representation (\ref{11}) is particularly suitable for an asymptotic expansion as $\beta\to 0$. Although for the phenomenon of Bose-Einstein condensation the low temperature regime is the pertinent one, it is clear for example from the references \cite{haug97-225-18,kirsten96} that the high-temperature expansion remains valid up to the condensation point and it can be used to determine the critical temperature at which condensation occurs.

The high-temperature expansion of (\ref{11}) is obtained by shifting the contour to the left. In that process we pick up contributions from poles of the spectral $\zeta$-function $\zeta(\alpha)$. The integral can then be computed by applying the Cauchy residue theorem.
The position of the poles and the corresponding residues are found by exploiting the intimate connection of $\zeta(\alpha)$ with its heat kernel
 \begin{equation}\label{12}
   K(t)={\sum_{i}}^{\prime}{\sum_{{\bf n}}}^{\prime}e^{-t\left(E_{{\bf n},i}-\mu_{c}\right)}\;.
 \end{equation}

The heat kernel in (\ref{12}) can be factorized as $K(t)=K_{H}(t)K_{\mathcal{N}}(t)$, where $K_{H}(t)$ is the heat kernel associated with the spectrum of the anisotropic harmonic oscillator and $K_{\mathcal{N}}(t)$ is the heat kernel for the Laplacian $\Delta_{\mathcal{N}}$ on the manifold $\mathcal{N}$ modified by the critical chemical potential which acts as a constant negative potential in (\ref{12}). The small-$t$ asymptotic expansion of (\ref{12}) which encodes the residues of $\zeta (\alpha )$ can be obtained from the one for $K_{H}(t)$ and $K_{\mathcal{N}}(t)$. The small-$t$ expansion for the harmonic oscillator part is trivially obtained because the heat-kernel can be computed in closed form by observing that the sums are simple infinite geometric series. For $K_{\mathcal{N}} (t)$, under the made assumptions on ${\mathcal{N}}$, the small-$t$ expansion is well known and reads \cite{gilkey95,mina53-17-158,mina49-1-242}
$$K_{\mathcal{N}} (t) \sim \frac 1 {(4\pi t)^{d/2}} \sum_{j=0,1/2,1,...}^\infty \mathscr{A}^{\mathcal{N}}_j t^j$$
with the heat-kernel coefficients $\mathscr{A}^{\mathcal{N}}_j$ of the modified Laplacian on ${\mathcal{N}}$.

Combining the expansions of each factor, in detail we have
\begin{equation}\label{13}
K(t)=\frac{1}{\omega_1\omega_2\omega_3(4\pi)^{\frac{d}{2}}}\sum_{k=0}^{\infty}\left(\sum_{l=0}^{[k/2]}C_{l}(\omega)\mathscr{A}^{\mathcal{N}}_{\frac{k}{2}-l}\right) t^{\frac{k-d-6}{2}}\;,
\end{equation}
where $[x]$ represents the integer part of $x$, $\mathscr{A}^{\mathcal{N}}_{l}$ are the heat kernel coefficients on the manifold $\mathcal{N}$ as given above, and the
$C_{l}(\omega)$, which come from the small-$t$ expansion of $K_{H}(t)$, have the form
\begin{equation}\label{14}
  C_{l}(\omega)=(-1)^{l}\sum_{n=0}^{l}\sum_{j=0}^{n}\frac{B_{j}B_{n-j}B_{l-n}}{j!(n-j)!(l-n)!}\omega_{1}^{j}\omega_{2}^{n-j}\omega_{3}^{l-n}\;,
\end{equation}
with $B_j$ denoting the Bernoulli numbers \cite{gradshtein07}. From the knowledge of the asymptotic expansion of the heat kernel (\ref{13}),
one can show that the rightmost poles of $\zeta(\alpha)$ are located at $\alpha_{k}=(d-k)/2+3$, with $k=0,\ldots,(d+5)$ \cite{gilkey95}. Their residues are given by
\begin{equation}\label{15}
  \textrm{Res}\,\zeta(\alpha_{k})=\frac{1}{\Gamma(\alpha_{k})}\sum_{l=0}^{[k/2]}C_{l}(\omega)\mathscr{A}^{\mathcal{N}}_{\frac{k}{2}-l}\;.
\end{equation}
Taking into account the first three rightmost poles of $\zeta(\alpha)$, we obtain the following asymptotic expansion
for the partition function valid for $\beta\to 0$
\begin{eqnarray}
  q&=&q_{0}+\beta^{-\left(\frac{d+6}{2}\right)}\textrm{Li}_{\frac{d+8}{2}}\left[e^{-\beta(\mu_{c}-\mu)}\right]\frac{\mathscr{A}^{\mathcal{N}}_{0}}{\omega_1\omega_2\omega_3}
  +\beta^{-\left(\frac{d+5}{2}\right)}\textrm{Li}_{\frac{d+7}{2}}\left[e^{-\beta(\mu_{c}-\mu)}\right]\frac{\mathscr{A}^{\mathcal{N}}_{1/2}}{\omega_1\omega_2\omega_3}\label{16}\\
  &+&\beta^{-\left(\frac{d+4}{2}\right)}\textrm{Li}_{\frac{d+6}{2}}\left[e^{-\beta(\mu_{c}-\mu)}\right]\frac{1}{\omega_1\omega_2\omega_3}\left(\mathscr{A}^{\mathcal{N}}_{1}
  +\frac{1}{2}\mathscr{A}^{\mathcal{N}}_{0}(\omega_1+\omega_2+\omega_3)\right)+O\left(\beta^{-\frac{d+3}{2}}\right)\;. \nonumber
\end{eqnarray}
The particle number $N$ is of specific importance in the analysis of Bose-Einstein condensation since it is used in order to define and compute the critical temperature.
It is well known that $N$ can be obtained from $q$ as
\begin{equation}\label{17}
  N=\frac{1}{\beta}\frac{\partial q}{\partial \mu}\;,
\end{equation}
with the derivative evaluated at fixed temperature and volume. The last remark, together with the result (\ref{16}), allows us to write
\begin{eqnarray}
  N&=&N_{0}+\beta^{-\left(\frac{d+6}{2}\right)}\textrm{Li}_{\frac{d+6}{2}}\left[e^{-\beta(\mu_{c}-\mu)}\right]\frac{\mathscr{A}^{\mathcal{N}}_{0}}{\omega_1\omega_2\omega_3}
  +\beta^{-\left(\frac{d+5}{2}\right)}\textrm{Li}_{\frac{d+5}{2}}\left[e^{-\beta(\mu_{c}-\mu)}\right]\frac{\mathscr{A}^{\mathcal{N}}_{1/2}}{\omega_1\omega_2\omega_3}\label{18}\\
  &+&\beta^{-\left(\frac{d+4}{2}\right)}\textrm{Li}_{\frac{d+4}{2}}\left[e^{-\beta(\mu_{c}-\mu)}\right]\frac{1}{\omega_1\omega_2\omega_3}\left(\mathscr{A}^{\mathcal{N}}_{1}
  +\frac{1}{2}\mathscr{A}^{\mathcal{N}}_{0}(\omega_1+\omega_2+\omega_3)\right)+O\left(\beta^{-\frac{d+3}{2}}\right)\;.\nonumber
\end{eqnarray}

The partition function $q$ also provides the energy of the system through the relation
\begin{equation}\label{18a}
  U=\left\{-\frac{\partial}{\partial\beta}+\frac{\mu}{\beta}\frac{\partial}{\partial\mu}\right\}q\;,
\end{equation}
 and it reads
 \begin{eqnarray}\label{18b}
   U&=&U_0+\left(\frac{d+6}{2}\right)\beta^{-\left(\frac{d+8}{2}\right)}\textrm{Li}_{\frac{d+8}{2}}\left[e^{-\beta(\mu_{c}-\mu)}\right]\frac{\mathscr{A}^{\mathcal{N}}_{0}}{\omega_1\omega_2\omega_3}
  +\left(\frac{d+5}{2}\right)\beta^{-\left(\frac{d+7}{2}\right)}\textrm{Li}_{\frac{d+7}{2}}\left[e^{-\beta(\mu_{c}-\mu)}\right]\frac{\mathscr{A}^{\mathcal{N}}_{1/2}}{\omega_1\omega_2\omega_3}\nonumber\\
  &+&\left(\frac{d+4}{2}\right)\beta^{-\left(\frac{d+6}{2}\right)}\textrm{Li}_{\frac{d+6}{2}}\left[e^{-\beta(\mu_{c}-\mu)}\right]\frac{1}{\omega_1\omega_2\omega_3}\left(\mathscr{A}^{\mathcal{N}}_{1}
  +\frac{1}{2}\mathscr{A}^{\mathcal{N}}_{0}(\omega_1+\omega_2+\omega_3)\right)\nonumber\\
  &+&\mu_{c}\beta^{-\left(\frac{d+6}{2}\right)}\textrm{Li}_{\frac{d+6}{2}}\left[e^{-\beta(\mu_{c}-\mu)}\right]\frac{\mathscr{A}^{\mathcal{N}}_{0}}{\omega_1\omega_2\omega_3}
  +\mu_{c}\beta^{-\left(\frac{d+5}{2}\right)}\textrm{Li}_{\frac{d+5}{2}}\left[e^{-\beta(\mu_{c}-\mu)}\right]\frac{\mathscr{A}^{\mathcal{N}}_{1/2}}{\omega_1\omega_2\omega_3}\nonumber\\
  &+&\mu_{c}\beta^{-\left(\frac{d+4}{2}\right)}\textrm{Li}_{\frac{d+4}{2}}\left[e^{-\beta(\mu_{c}-\mu)}\right]\frac{1}{\omega_1\omega_2\omega_3}\left(\mathscr{A}^{\mathcal{N}}_{1}
  +\frac{1}{2}\mathscr{A}^{\mathcal{N}}_{0}(\omega_1+\omega_2+\omega_3)\right)
  +O\left(\beta^{-\frac{d+3}{2}}\right)\;.
 \end{eqnarray}
This expression for the energy will be used, in what follows, in order to compute the specific heat of the Bose gas.

\section{Critical Temperature and Specific Heat}

The critical temperature $T_{c}=1/\beta_{c}$, at which the condensate starts to appear, is obtained from the particle number $N$ in (\ref{18}) by setting $N_{0}=0$.
When the temperature of the system approaches the critical temperature we have $\mu\sim\mu_{c}$ and, hence, we can use the Taylor expansion, valid for $n>2$,
\begin{equation}\label{19}
\textrm{Li}_{n}(e^{-x})=\zeta_{R}(n)-x\zeta_{R}(n-1)+O(x^{2})\;.
\end{equation}
In these circumstances, the critical temperature is approximately defined according to the relation
\begin{eqnarray}\label{20}
  N&=&\beta^{-\left(\frac{d+6}{2}\right)}\zeta_{R}\left(\frac{d+6}{2}\right)\frac{\mathscr{A}^{\mathcal{N}}_{0}}{\omega_1\omega_2\omega_3}
  +\beta^{-\left(\frac{d+5}{2}\right)}\zeta_{R}\left(\frac{d+5}{2}\right)\frac{\mathscr{A}^{\mathcal{N}}_{1/2}}{\omega_1\omega_2\omega_3}\nonumber\\
  &+&\beta^{-\left(\frac{d+4}{2}\right)}\zeta_{R}\left(\frac{d+4}{2}\right)\frac{1}{\omega_1\omega_2\omega_3}\left(\mathscr{A}^{\mathcal{N}}_{1}
  +\frac{1}{2}\mathscr{A}^{\mathcal{N}}_{0}(\omega_1+\omega_2+\omega_3)\right)+O\left(\beta^{-\frac{d+3}{2}}\right)\;,
\end{eqnarray}
and it can be found to be
\begin{equation}\label{21}
  T_{c}=T_{0}\left\{1-\frac{2\zeta_{R}\left(\frac{d+5}{2}\right)\mathscr{A}^{\mathcal{N}}_{1/2}}{(d+6)(\omega_1\omega_2\omega_3)^{\frac{1}{d+6}}
  \zeta_{R}\left(\frac{d+6}{2}\right)^{\frac{d+5}{d+6}}\left(\mathscr{A}_{0}^{\mathcal{N}}\right)^{\frac{d+5}{d+6}}}N^{-\frac{1}{d+6}}\right\}\;,
\end{equation}
if $\partial\mathcal{N}\neq 0$ and
\begin{equation}\label{22}
   T_{c}=T_{0}\left\{1-\frac{2\zeta_{R}\left(\frac{d+4}{2}\right)\left[\mathscr{A}^{\mathcal{N}}_{1}+\frac{1}{2}\mathscr{A}^{\mathcal{N}}_{0}(\omega_1+\omega_2+\omega_3)\right]}
   {(d+6)(\omega_1\omega_2\omega_3)^{\frac{2}{d+6}}
  \zeta_{R}\left(\frac{d+6}{2}\right)^{\frac{d+4}{d+6}}\left(\mathscr{A}_{0}^{\mathcal{N}}\right)^{\frac{d+4}{d+6}}}N^{-\frac{2}{d+6}}\right\}\;,
\end{equation}
when $\partial\mathcal{N}=0$. We would like to point out that in (\ref{21}) and (\ref{22}) we have defined, for brevity,
\begin{equation}\label{23}
  T_{0}=(\omega_1\omega_2\omega_3)^{\frac{2}{d+6}}\left[\frac{N}{\zeta_{R}\left(\frac{d+6}{2}\right)\mathscr{A}_{0}^{\mathcal{N}}}\right]^{\frac{2}{d+6}}\;.
\end{equation}
Equation (\ref{22}) shows in detail how the critical temperature correction due to the finite number of particles depends on properties of the extra dimensions and, of course, the frequencies of the harmonic oscillator potential.
In some detail, the leading heat kernel coefficients are well known \cite{gilkey95} and we have that $\mathscr{A}^{\mathcal N}_0$ equals the volume of $\mathcal N$, $\mathscr{A}^{\mathcal N}_{1/2}$ is proportional to the volume of the boundary of $\mathcal N$, and higher order terms involve curvature tensors of the manifold $\mathcal N$ and its boundary. Thus a measurement of the critical temperature as a function of the finite particle number $N$ could reveal properties of the extra dimension like its number and size.

A further interesting thermodynamical quantity associated with the system is the specific heat. It is derived from the energy according to the relation
\begin{equation}\label{24}
  C=\frac{\partial U}{\partial T}\;,
\end{equation}
where the derivative is understood as performed by keeping both the number of particles and the volume constant.
From  (\ref{18b}) and (\ref{24}), near the critical temperature $T_c$, one has
\begin{eqnarray}\label{25}
  C&=&\frac{(d+8)(d+6)}{4\omega_1\omega_2\omega_3}\beta^{-\left(\frac{d+6}{2}\right)}\zeta_{R}\left(\frac{d+8}{2}\right)\mathscr{A}^{\mathcal{N}}_{0}
  +\frac{(d+7)(d+5)}{4\omega_1\omega_2\omega_3}\beta^{-\left(\frac{d+5}{2}\right)}\zeta_{R}\left(\frac{d+7}{2}\right)\mathscr{A}^{\mathcal{N}}_{1/2}\nonumber\\
& &  +O\left(\beta^{-\frac d 2 -2}\right)\;,
\end{eqnarray}
when the manifold $\mathcal{N}$ satisfies the property $\partial\mathcal{N}\neq 0$, and
\begin{eqnarray}\label{26}
  C&=&\frac{(d+8)(d+6)}{4\omega_1\omega_2\omega_3}\beta^{-\left(\frac{d+6}{2}\right)}\zeta_{R}\left(\frac{d+8}{2}\right)\mathscr{A}^{\mathcal{N}}_{0}\nonumber\\
  &+&\frac{(d+6)(d+4)}{4\omega_1\omega_2\omega_3}\beta^{-\left(\frac{d+4}{2}\right)}\zeta_{R}\left(\frac{d+6}{2}\right)\left(\mathscr{A}^{\mathcal{N}}_{1}
  +\frac{1}{2}\mathscr{A}^{\mathcal{N}}_{0}(\omega_1+\omega_2+\omega_3)\right)\nonumber\\
  &-&\frac{(\mu_c-\mu)^{2}(d+6)^{2}}{4g_{0}(\omega_1\omega_2\omega_3)^{2}}\beta^{-d-4}\zeta_{R}\left(\frac{d+6}{2}\right)^{2}\left(\mathscr{A}^{\mathcal{N}}_{0}\right)^{2}
  +O\left(\beta^{-\frac d 2 -1}\right)\;,
\end{eqnarray}
for $\partial\mathcal{N}=0$. The results that we have obtained for the critical temperature (\ref{21})-(\ref{22}) and for the specific heat (\ref{25})-(\ref{26}) are
very general and hold for an arbitrary smooth, compact manifold $\mathcal{N}$. More explicit formulas can be obtained once the manifold $\mathcal{N}$ has been specified.
In this case only the knowledge of the first few heat kernel coefficients is necessary, which are well known for a wide variety of manifolds with and without boundary \cite{kirsten01,vassilevich03}.

\section{Final Remarks}

In this work we have presented an analysis of Bose-Einstein condensation on product manifolds by utilizing $\zeta$-function
regularization techniques. The method proves to be very efficient and close to the condensation temperature the general results for the critical temperature and the specific heat depend only on the first few readily available heat kernel coefficients associated with the base manifold $\mathcal{N}$. It is important to notice that the dependence of the critical temperature on the finite number $N$ of particles reveals geometrical and topological properties of the extra Kaluza-Klein dimensions; see equations (\ref{21}) and (\ref{22}).

It would be of particular interest to study in more detail the cases in which the manifold $\mathcal{N}$ is either a $d$-dimensional sphere or a $d$-dimensional torus.
In these situations the spectrum $\lambda_{i}$ is explicitly known and for the sphere one would be led to deal with Barnes $\zeta$-functions, while for the torus Epstein $\zeta$-functions would be the
relevant objects. We plan to investigate these cases in a future work. Furthermore, similar calculations should be done for string inspired models such that future experiments could possibly determine features of extra dimensions in these models too.\\[.3cm]
\noindent
{\bf Acknowledgments:} KK acknowledges support by the National Science Foundation Grant PHY-0757791.


\begin{thebibliography}{14}
\expandafter\ifx\csname natexlab\endcsname\relax\def\natexlab#1{#1}\fi
\expandafter\ifx\csname bibnamefont\endcsname\relax
  \def\bibnamefont#1{#1}\fi
\expandafter\ifx\csname bibfnamefont\endcsname\relax
  \def\bibfnamefont#1{#1}\fi
\expandafter\ifx\csname citenamefont\endcsname\relax
  \def\citenamefont#1{#1}\fi
\expandafter\ifx\csname url\endcsname\relax
  \def\url#1{\texttt{#1}}\fi
\expandafter\ifx\csname urlprefix\endcsname\relax\def\urlprefix{URL }\fi
\providecommand{\bibinfo}[2]{#2}
\providecommand{\eprint}[2][]{\url{#2}}

\bibitem{altaie78} M. B. Al'taie, \emph{J. Phys.} A {\bf 11} 1603 (1978).

\bibitem[{\citenamefont{Anderson et~al.}(1995)\citenamefont{Anderson, Ensher,
  Matthews, Wieman, and Cornell}}]{ande95-269-198}
\bibinfo{author}{\bibfnamefont{M.}~\bibnamefont{Anderson}},
  \bibinfo{author}{\bibfnamefont{J.}~\bibnamefont{Ensher}},
  \bibinfo{author}{\bibfnamefont{M.}~\bibnamefont{Matthews}},
  \bibinfo{author}{\bibfnamefont{C.}~\bibnamefont{Wieman}}, \bibnamefont{and}
  \bibinfo{author}{\bibfnamefont{E.}~\bibnamefont{Cornell}},
  \bibinfo{journal}{Science} \textbf{\bibinfo{volume}{269}},
  \bibinfo{pages}{198} (\bibinfo{year}{1995}).

\bibitem{bagnato87} V. Bagnato, D. E. Pritchard, and D. Kleppner, {\emph Phys. Rev.} A {\bf 35} 4354 (1987).

\bibitem[{\citenamefont{Bordag et~al.}(2009)\citenamefont{Bordag,
  Klimchitskaya, Mohideen, and Mostepanenko}}]{bord09b}
\bibinfo{author}{\bibfnamefont{M.}~\bibnamefont{Bordag}},
  \bibinfo{author}{\bibfnamefont{G.}~\bibnamefont{Klimchitskaya}},
  \bibinfo{author}{\bibfnamefont{U.}~\bibnamefont{Mohideen}}, \bibnamefont{and}
  \bibinfo{author}{\bibfnamefont{V.}~\bibnamefont{Mostepanenko}},
  \emph{\bibinfo{title}{Advances in the Casimir effect}}
  (\bibinfo{publisher}{Oxford Science Publications}, \bibinfo{year}{2009}).

\bibitem{bose} S. N. Bose, \emph{Z. Phys.} {\bf 26} 178 (1924).

\bibitem[{\citenamefont{Bradley et~al.}(1995)\citenamefont{Bradley, Sackett,
  Tollett, and Hulet}}]{brad95-75-1687}
\bibinfo{author}{\bibfnamefont{C.}~\bibnamefont{Bradley}},
  \bibinfo{author}{\bibfnamefont{C.}~\bibnamefont{Sackett}},
  \bibinfo{author}{\bibfnamefont{J.}~\bibnamefont{Tollett}}, \bibnamefont{and}
  \bibinfo{author}{\bibfnamefont{R.}~\bibnamefont{Hulet}},
  \bibinfo{journal}{Phys. Rev. Lett.} \textbf{\bibinfo{volume}{75}},
  \bibinfo{pages}{1687} (\bibinfo{year}{1995}).

\bibitem{bytsenko92} A. A. Bytsenko, L. Vanzo, and S. Zerbini, \emph{Phys. Lett.} B {\bf 291} 26 (1992).

\bibitem[{\citenamefont{Cheng}(2008)}]{chen08-668-72}
\bibinfo{author}{\bibfnamefont{H.}~\bibnamefont{Cheng}},
  \bibinfo{journal}{Phys. Lett.} \textbf{\bibinfo{volume}{B668}},
  \bibinfo{pages}{72} (\bibinfo{year}{2008}).

\bibitem[{\citenamefont{Davis et~al.}(1995)\citenamefont{Davis, Mewes, Andrews,
  van Druten, Durfee, Kurn, and Ketterle}}]{davi95-75-3969}
\bibinfo{author}{\bibfnamefont{K.}~\bibnamefont{Davis}},
  \bibinfo{author}{\bibfnamefont{M.-O.} \bibnamefont{Mewes}},
  \bibinfo{author}{\bibfnamefont{M.}~\bibnamefont{Andrews}},
  \bibinfo{author}{\bibfnamefont{N.}~\bibnamefont{van Druten}},
  \bibinfo{author}{\bibfnamefont{D.}~\bibnamefont{Durfee}},
  \bibinfo{author}{\bibfnamefont{D.}~\bibnamefont{Kurn}}, \bibnamefont{and}
  \bibinfo{author}{\bibfnamefont{W.}~\bibnamefont{Ketterle}},
  \bibinfo{journal}{Phys. Rev. Lett.} \textbf{\bibinfo{volume}{75}},
  \bibinfo{pages}{3969} (\bibinfo{year}{1995}).

\bibitem{degroot50} S. R. de Groot, G. J. Hooyman, and C. A. ten Seldam, \emph{Proc. R. Soc. London} A {\bf 203} 266 (1950).

\bibitem{dowker78} J. S. Dowker, and G. Kennedy, \emph{J. Phys.} A {\bf 11} 895 (1978).

\bibitem[{\citenamefont{Dowker and Schofield}(1989)}]{dowk89-327-267}
\bibinfo{author}{\bibfnamefont{J.}~\bibnamefont{Dowker}} \bibnamefont{and}
  \bibinfo{author}{\bibfnamefont{J.}~\bibnamefont{Schofield}},
  \bibinfo{journal}{Nucl. Phys.} \textbf{\bibinfo{volume}{B327}},
  \bibinfo{pages}{267} (\bibinfo{year}{1989}).

\bibitem{einstein} A. Einstein, \emph{Sitzungsber. Preus. Akad. Wiss.} {\bf 22} 261 (1924).

\bibitem{elizalde94} E. Elizalde E, S. D. Odintsov, A. Romeo, A. A. Bytsenko, and S. Zerbini, Zeta Regularization Techniques with Applications,
(World Scientific, Singapore) (1994).

\bibitem{elizalde} E. Elizalde, Ten Physical Applications of the Spectral Zeta Function,
(Springer-Verlag, Berlin) (1995).

\bibitem{elizalde95} E. Elizalde, K. Kirsten, and S. Zerbini, \emph{J. Phys.} A {\bf 28} 617 (1995).

\bibitem[{\citenamefont{Frank et~al.}(2007)\citenamefont{Frank, Turan, and
  Ziegler}}]{fran07-76-015008}
\bibinfo{author}{\bibfnamefont{M.}~\bibnamefont{Frank}},
  \bibinfo{author}{\bibfnamefont{I.}~\bibnamefont{Turan}}, \bibnamefont{and}
  \bibinfo{author}{\bibfnamefont{L.}~\bibnamefont{Ziegler}},
  \bibinfo{journal}{Phys. Rev.} \textbf{\bibinfo{volume}{D76}},
  \bibinfo{pages}{015008} (\bibinfo{year}{2007}).

\bibitem{gilkey95} P.B. Gilkey, Invariance Theory the Heat Equation and the Atiyah-Singer Index Theorem,
 (Boca raton: CRC Press) (1995).

\bibitem{gradshtein07}  I. S. Gradshtein, and I. M. Ryzhik,
Table of Integrals, Series and Products, Eds. A. Jeffrey and D.
Zwillinger (Oxford: Academic) (2007).

\bibitem{greenspoon75} S. Greenspoon, and R. K. Pathria, \emph{Phys. Rev.} A {\bf 11} 1080 (1975).

\bibitem{grossmann95} S. Grossmann, and M. Holthaus, \emph{Phys. Lett.} A {\bf 208} 188 (1995).

\bibitem{grossmann97} S. Grossmann, and M. Holthaus, \emph{Optics Express} {\bf 1} 262 (1997).

\bibitem[{\citenamefont{Haugerud et~al.}(1997)\citenamefont{Haugerud, Haugset,
  and Ravndal}}]{haug97-225-18}
\bibinfo{author}{\bibfnamefont{H.}~\bibnamefont{Haugerud}},
  \bibinfo{author}{\bibfnamefont{T.}~\bibnamefont{Haugset}}, \bibnamefont{and}
  \bibinfo{author}{\bibfnamefont{F.}~\bibnamefont{Ravndal}},
  \bibinfo{journal}{Phys. Lett.} \textbf{\bibinfo{volume}{A225}},
  \bibinfo{pages}{18} (\bibinfo{year}{1997}).

\bibitem{haug97-55-2922} T. Haugset, H. Haugerud, and J. O. Anderson, \emph{Phys. Rev.} A {\bf 55} 2922 (1997).

\bibitem[{\citenamefont{Hofmann et~al.}(2004)\citenamefont{Hofmann,
  Poppenhaeger, Hossenfelder, and Bleicher}}]{hofm04-582-1}
\bibinfo{author}{\bibfnamefont{S.}~\bibnamefont{Hofmann}},
  \bibinfo{author}{\bibfnamefont{K.}~\bibnamefont{Poppenhaeger}},
  \bibinfo{author}{\bibfnamefont{S.}~\bibnamefont{Hossenfelder}},
  \bibnamefont{and} \bibinfo{author}{\bibfnamefont{M.}~\bibnamefont{Bleicher}},
  \bibinfo{journal}{Phys. Lett. B} \textbf{\bibinfo{volume}{582}},
  \bibinfo{pages}{1} (\bibinfo{year}{2004}).

\bibitem[{\citenamefont{Inouye et~al.}(1998)\citenamefont{Inouye, Andrews,
  Stenger, Miesner, Stamper-Kurn, and Ketterle}}]{inou98-392-151}
\bibinfo{author}{\bibfnamefont{S.}~\bibnamefont{Inouye}},
  \bibinfo{author}{\bibfnamefont{M.}~\bibnamefont{Andrews}},
  \bibinfo{author}{\bibfnamefont{J.}~\bibnamefont{Stenger}},
  \bibinfo{author}{\bibfnamefont{H.-J.} \bibnamefont{Miesner}},
  \bibinfo{author}{\bibfnamefont{D.}~\bibnamefont{Stamper-Kurn}},
  \bibnamefont{and} \bibinfo{author}{\bibfnamefont{W.}~\bibnamefont{Ketterle}},
  \bibinfo{journal}{Nature} \textbf{\bibinfo{volume}{392}},
  \bibinfo{pages}{151} (\bibinfo{year}{1998}).

\bibitem{kirsten96} K. Kirsten, and D. J. Toms, \emph{Phys. Rev.} A {\bf 54} 4188 (1996).

\bibitem{kirsten96a} K. Kirsten, and D. J. Toms, \emph{Phys. Lett.} A {\bf 222} 148 (1996).

\bibitem{kirsten98} K. Kirsten, and D. J. Toms, \emph{Phys. Lett.} A {\bf 243} 137 (1998).

\bibitem{kirsten99} K. Kirsten, and D. J. Toms, \emph{Phys. Rev.} E {\bf 59} 158 (1999).

\bibitem[{\citenamefont{Kirsten}(1991)}]{kirs91-8-2239}
\bibinfo{author}{\bibfnamefont{K.}~\bibnamefont{Kirsten}},
  \bibinfo{journal}{Class. Quantum Grav.} \textbf{\bibinfo{volume}{8}},
  \bibinfo{pages}{2239} (\bibinfo{year}{1991}).

\bibitem{kirsten01}
K. Kirsten,
Spectral Functions in Mathematics and Physics,
(Boca Raton: CRC Press) (2001).

\bibitem{landau} L. D. Landau, and E. M. Lifshitz, Statistical Physics, (Pergamon, London) (1969).

\bibitem[{\citenamefont{Minakshisundaram}(1953)}]{mina53-17-158}
\bibinfo{author}{\bibfnamefont{S.}~\bibnamefont{Minakshisundaram}},
  \bibinfo{journal}{J. Indian Math. Soc.} \textbf{\bibinfo{volume}{17}},
  \bibinfo{pages}{158} (\bibinfo{year}{1953}).

\bibitem[{\citenamefont{Minakshisundaram and Pleijel}(1949)}]{mina49-1-242}
\bibinfo{author}{\bibfnamefont{S.}~\bibnamefont{Minakshisundaram}}
  \bibnamefont{and} \bibinfo{author}{\bibfnamefont{A.}~\bibnamefont{Pleijel}},
  \bibinfo{journal}{Can. J. Math.} \textbf{\bibinfo{volume}{1}},
  \bibinfo{pages}{242} (\bibinfo{year}{1949}).

\bibitem{pajkowski77} H. R. Pajkowski, and R. K. Pathria, \emph{J. Phys.} A {\bf 10} 561 (1977).

\bibitem{parker91} L. Parker, and Y. Zhang, \emph{Phys. Rev.} D {\bf 44} 2421 (1991).

\bibitem{pathria72} R. K. Pathria, \emph{Phys. Rev.} A {\bf 5} 1451 (1972).

\bibitem{shiraishi87} K. Shiraishi, \emph{Prog. Theor. Phys.} {\bf 77} 975 (1987).

\bibitem{singh84} S. Singh, and R. K. Pathria, \emph{J. Phys.} A {\bf 17} 2983 (1984).

\bibitem{smith96} J. D. Smith, and D. J. Toms, \emph{Phys. Rev.} D {\bf 53} 5771 (1996).

\bibitem{toms92} D. J. Toms, \emph{Phys. Rev. Lett.} {\bf 69} 1152 (1992).

\bibitem{toms93} D. J. Toms, \emph{Phys. Rev.} D {\bf 47} 2483 (1993).

\bibitem{vassilevich03}  D. V. Vassilevich, \emph{Phys. Rep.} {\bf 388} 279 (2003).


\end{thebibliography}
\end{document}